\documentstyle[preprint,epsfig,eqsecnum,aps]{revtex}
\begin{document}
\draft

\title{New features of collective motion of intrinsic degrees of freedom.
Toward a possible way to classify the intrinsic states}

\author{A. A. Raduta$^{a),b)}$, L. Pacearescu$^{a),c)}$, V. Baran$^{a)}$}

\address{$^{a)}$Institute of Physics and Nuclear Engineering, Bucharest,
POBox MG6, Romania}

\address{$^{b)}$Department of Theoretical Physics and Mathematics,
Faculty of Physics, Bucharest University, POBox MG11, Romania}

\address{$^{c)}$Institut fuer Theoretische Physik der Universitaet Tuebingen,
Auf der Morgenstelle 14, Germany}

\maketitle
\date{today}
\abstract{Three exactly solvable Hamiltonians of complex structure, are
studied in the framework of a semi-classical approach. The quantized
trajectories for intrinsic coordinates correspond to energies 
which may be classified in collective bands. For two of the chosen Hamiltonians
the symmetry SU(2)$\otimes$SU(2) is the appropriate one to classify the eigenvalues
in the laboratory frame. Connections of results presented here with the molecular spectrum and
Moszkowski model are pointed out. The present approach suggests that
the intrinsic states, which in  standard formalisms are heading
rotational bands, are forming themselves "rotational" bands,  the rotations
being performed in a fictitious boson space.
\section{Introduction}
Collective motion in nuclei has constantly been an exciting subject for both 
experimentalists and theoreticians. Some phenomenological models 
have been successfully in interpreting the data for energies and
transition probabilities for certain ranges of atomic mass.
The long series of publications has been opened by the liquid drop model
\cite{Bohr}  which
was later on improved by including anharmonicities, deformed
equilibrium shapes or by extending the 
symmetries for the intrinsic motion
\cite{Faes,Gneus,Hess,Grei0,Lie,Lip,Rad11,Arim}. Some phenomenological models
interpret the properties of collective states of positive parity in
terms of quadrupole degrees of freedom. Interacting boson model 
(IBA)\cite{Arim} points out the fact that the monopole degrees of freedom 
should be included which results in obtaining a conceptually new formalism for
the description of nuclear dynamics.
All phenomenological models define notions like rotational bands,
equilibrium shapes, nuclear phases based on the behavior of the system
in the intrinsic frame.
The coherent state model (CSM) exploits the classical features of
collective states of high angular momentum and provides
a realistic description for deformed and transitional region in terms of
angular momentum projected states from coherent states\cite{Rad11}. 

Some microscopic theories were developed by paralleling the aims of the
phenomenological models. Indeed, the nuclear properties are described in
terms of individual degrees of freedom. In this sense to define the optimal
collective coordinates for a many body system, was always a
central field of activity. Microscopic formalisms use approximations
like Bogoliubov-Hartree-Fok, Random Phase Approximation (RPA), boson
expansion technic, higher order  RPA, extended shell model, etc. \cite{Ring},
which need to be tested and define the circumstances under which they 
work or not. For such purposes very often one uses completely solvable models
where the results of various approaches are compared with the exact result.
Most known schematic and solvable models are those of Moszkowski \cite{Mosz}, 
Lipkin-Meshkov \cite{Lipk} and One Level Pairing Hamiltonian 
\cite{Maf,Ri}. Also phenomenological
models proposed by Wilets and Jean\cite{Wil}, Davydov and Filippov\cite{Dav} 
are even nowadays used to get a reference framework which allows for an
interpretation of the results obtained with more sophisticated models.  

In this paper we address the question whether the motion of intrinsic
collective coordinates can be described by the irreducible
representation of a symmetry group. The chosen model Hamiltonians have
a complex structure being of fourth order in quadrupole bosons. They
have however a common feature that all of them are, in a semi-classical
picture, completely solvable and therefore analytical expressions for
energies are obtained.

The model Hamiltonians are explicitly treated according to the
following plan:
In Section 2 a fourth order Hamiltonoian consisting in a quadratic
quadrupole boson term plus a fourth order term
which is  the square of the second order boson invariant. Two sets of
parameters are considered, one yielding, after dequantization, an
effective potential with one minimum and  an other one which produces
a pocket potential. Solutions are presented for both cases. Energies
were organized in bands like standard bands are in the laboratory frame.
The difference is that here the angular momentum is defined by an
${\cal R}_3$  group acting in a fictitious space.
In Section 3 the two body boson interaction is of a multipole-multipole type
with the multipolarity $2^k,k=0,2,4$.
In the intrinsic frame the classical Hamiltonian has energies which
after quantization might be classified by L,M and therefore the
corresponding states are SU(2) states. This pictures implies that in
the laboratory frame the symmetry SU(2)$\otimes$SU(2) might be used.
In Section 4, the two body boson interaction is of a 
quadrupole-quadrupole type. 
Neglecting the correlations with the excited collective
bands the classical Hamiltonian is identical with those proposed by Moszkowski
\cite{Mosz} and therefore a simple expression for the corresponding
spectrum is possible.
Final conclusions are drawn in Section 5.

\label{sec:level1}
\section{The first example of an exactly solvable Hamiltonian}
\label{sec:level2}
In this Section we analyze the classical behavior of a fourth order
boson Hamiltonian with a particular structure. The third order term is
missing and the fourth order one depends exclusively on quadrupole
coordinates but on their conjugate momenta:
\begin{equation}
H_1=A_1(b^{\dagger}_2b_2)_0+A_2\left[(b^{\dagger}_2b^{\dagger}_2)_0+
(b_2b_2)_0\right]+A_4{\hat P}^2.
\label{H1}
\end{equation}
Here $b^{\dagger}_{2\mu}$ ($b_{2\mu}$) denotes the creation
(annihilation) quadrupole boson operator while ${\hat P}$ is the second
order boson invariant:
\begin{equation}
{\hat P}=\frac{1}{2}\sum_{\mu}(b^{\dagger}_{2\mu}+(-)^{\mu}b_{2-\mu})
(b^{\dagger}_{2-\mu}+(-)^{\mu}b_{2\mu})(-)^\mu.
\label{P}
\end{equation}
We consider the variational equation 
\begin{equation} 
\delta \int_{0}^{t} \langle \psi \mid H_1 - i \hbar  
\frac{\partial}{\partial t'} \mid \psi \rangle dt' = 0 
\label{varpr}
\end{equation} 
\noindent 
where $\mid \Psi \rangle$ denotes the following coherent state
\begin{equation} 
\mid \psi \rangle = exp[ z_{0}b_{20}^{+} - z_{0}^{*}b_{20} + 
z_{2}( b_{22}^{+} + b_{2-2}^{+} ) - z_{2}^{*}( b_{22} + b_{2-2} )]
\mid 0 \rangle
\label{Psi} 
\end{equation} 
The vacuum state for the quadrupole boson operators $b_{2\mu}$ is denoted
by  $\mid 0 \rangle$; $z_{0}$ and $z_{2}$ are complex functions of time
and play the role of classical coordinates. To simplify the notations,
hereafter the quadrupole boson operators will be denoted by $b_{\mu}^{+}$
omitting the index specifying the angular momentum carried by the boson
operators. 

The coordinate transformation:
\begin{eqnarray} 
q_{i}& =& 2^{\frac{k+2}{4}}Re(z_{k}),\nonumber\\
p_{i}& =& 2^{\frac{k+2}{4}}Im(z_{k}),~~k = 0, 2,~~ i=\frac{k+2}{4},
\label{qandp}
\end{eqnarray}
brings the classical equation of motion to a canonical form:
\begin{eqnarray}  
\frac{\partial~ {\cal H}_1}{\partial~q_{k}}&=&-\dot{p}_{k},
\nonumber\\
\frac{\partial~ {\cal H}_1}{\partial~p_{k}}&=&\dot{q}_{k}.
\label{eom1}
\end{eqnarray}
Here ${\cal H}_1$ stands for the average value of the chosen Hamiltonian 
on the coherent state $\mid \Psi \rangle$ while  
"dot" denotes the time derivative operation.
\begin{equation} 
{\cal H}_1 \equiv \langle \psi \mid H_1 \mid \psi \rangle =
\frac{A'}{2}( p_{1}^{2} + p_{2}^{2}) +
\frac{A}{2}( q_{1}^{2} + q_{2}^{2} ) + 
D ( q_{1}^{2} + q_{2}^{2} )^{2}.
\label{Hrond1}  
\end{equation} 
The coefficients involved in the expression of ${\cal H}_1$ are related
to those defining the model Hamiltonian by the following equations:
\begin{eqnarray}
A&=&\frac{1}{\sqrt{5}}(A_1+2A_2)+14A_4,\nonumber\\
A'&=&\frac{1}{\sqrt{5}}(A_1-2A_2),\nonumber\\
D&=&4A_4.
\end{eqnarray}
For what follow, it is useful to introduce the polar coordinates:
\begin{equation} 
q_{1} = r \cos \theta~~,~~q_{2} = r \sin \theta .
\label{q1andq2}
\end{equation}
In terms of the new coordinates the classical energy function has the expression:
\begin{equation} 
{\cal H}_1= \frac{\hbar^2}{2 A'}(\dot{r}^{2}+r^2\dot{\theta}^2) +
\frac{A}{2}r^2+\frac{D}{4}r^4.
\label{H1polarcoord}
\end{equation}
From the above equations we see that the classical system associated to the
boson Hamiltonian is a particle which moves in a plane due to the force
determined by a potential invariant with respect to rotations 
around
an axes perpendicular to the motion plane:
\begin{equation} 
V(r)=\frac{A}{2}r^2+\frac{D}{4}r^4.
\label{Vofr}
\end{equation}
The potential energy is plotted in Fig.1 for three sets of parameters (A,D).
The parameters were chosen so that three distinct situations are pointed
out, namely when the equilibrium shape is spherical, deformed and meta-stable.
Using the equations of motion (\ref{eom1}) one can prove that
\begin{equation}
\dot{\cal L}_3=0~~,~~\dot{\cal H}_1=0.
\label{cmot}
\end{equation}
where ${\cal L}_{3}$ is defined by the following expression:
\begin{equation}
{\cal L}_{3} \equiv \frac{\hbar}{2}
(q_{1} p_{2} - q_{2} p_{1}) =
\frac{\hbar^2}{A'} r^2 \dot{\theta}.
\label{L3}  
\end{equation} 
Let us define another two classical functions on the phase space, spanned 
by the coordinates ($q_{1},p_{1},q_{2},p_{2}$)
\begin{eqnarray}
{\cal L}_{1}& = &\frac{\hbar}{4}
((q_{1}^2 + p_{1}^{2} - q_{2}^{2}- p_{2}^{2}) 
\nonumber\\
{\cal L}_{2}&=&\frac{\hbar}{2}(q_{1} q_{2} + p_{1}p_{2}) .
\label{L1andL2}
\end{eqnarray}
Given two function $f_{1}$ and $f_{2}$, defined on the phase space, their
Poisson bracket is defined as:
\begin{equation}  
\{ f_{1}, f_{2} \} = \sum_{k=1,2} \left[ \frac{\partial~f_{1}}
{\partial~q_{k}}\frac{\partial~f_{2}}{\partial~p_{k}}-
\frac{\partial~f_{1}}{\partial~p_{k}}
\frac{\partial~f_{2}}{\partial~q_{k}}\right] .
\end{equation}
The classical functions ${\cal L}_{k}$ obey the following equations:
\begin{eqnarray}
\{{\cal L}_{1},{\cal L}_{2}\}&=&\hbar{\cal L}_{3},\nonumber\\
\{{\cal L}_{2},{\cal L}_{3}\}&=&\hbar{\cal L}_{1},\nonumber\\
\{{\cal L}_{3},{\cal L}_{1}\}&=&\hbar{\cal L}_{2}.
\label{L1L2L3}
\end{eqnarray}
In virtue of Eq.(\ref{L1L2L3}) the set of functions ${\cal L}_{k}$ with the 
Poisson brackets as multiplication
operation, form a classical  $SU_c(2)$ algebra. Moreover they could be obtained
by averaging on  $\mid \Psi \rangle$, the generators of a boson $SU_b(2)$ 
algebra
defined with the boson operators $b_{0}^{+},b_{\pm 2}^{+}$:
\begin{eqnarray}
 {\cal L}_{k} &=& \langle \psi \mid {\hat L}_{k} \mid \psi \rangle; k=1,2,3, 
\nonumber\\
 {\hat L}_{1}& = &\frac{\hbar}{4}\left[2 b_{0}^{\dagger}b_{0} - ( b_{2}^{\dagger}
+ 
b_{-2}^{\dagger} )( b_{2} + b_{-2} )\right],\nonumber \\
{\hat L}_{2}& = &\frac{\hbar}{2 \sqrt{2}}\left[b_{0}^{\dagger}(b_{2} + b_{-2})+ 
(b_{2}^{\dagger} + b_{-2}^{\dagger})b_{0}\right],\nonumber\\ 
{\hat L}_{3}&=&\frac{\hbar}{2 \sqrt{2} i}\left[b_{0}^{\dagger}(b_{2} + b_{-2})-
 (b_{2}^{\dagger} + b_{-2}^{\dagger})b_{0}\right].
\label{Lboz}
\end{eqnarray}
The first equation (\ref{Lboz}) and the correspondence between commutators and Poisson brackets
\begin{equation}
[,]  \rightarrow \frac{1}{i}\{,\},
\label{parant}
\end{equation}
define a homeomorphism of the boson and classical algebras generated by
$ \{ {\hat L}_{k} \}_{k=1,2,3}$  and $\{{\cal L}_{k}\}_{k=1,2,3}$ respectively.
Note that the boson $SU_b(2)$ algebra does not describe the rotations
in the real configuration space but in a fictitious space.
The conservation law expressed by (\ref{cmot}) is determined by the invariance against
rotation around the 3-rd axis in the fictitious space mentioned above:
\begin{equation}
[H_1,\hat{L}_{3}] = 0 .
\label{comH1L3}
\end{equation}
Since the classical system is characterized by two degrees of freedom and, on 
the other hand, there are two constants of motion
\begin{equation}
{\cal H}_1 = E~~ ,~~{\cal L}_{3} = \frac{L}{2} ,
\label{H1eqE}
\end{equation}
the equations of motion are exactly solvable. Indeed, inserting the constants
 of motion in Eq. (\ref{H1polarcoord}) the resulting differential equation
\begin{equation} 
\frac{\hbar^{2} \dot{r}^{2}}{2 A'} + V_{eff}(L;r)= E,
\label{toten}
\end{equation}
with
\begin{equation} 
V_{eff}(L;r)= \frac{A' L^{2}}{2{r}^{2}}+\frac{A}{2}r^2+\frac{D}{4}r^4,
\label{Veff}
\end{equation}
provides the time variable as function of $x=r^{2}$ by
\begin{equation} 
t = \frac{ \hbar}{\sqrt{-2A'D}} \int_{x_{0}}^{x}
 \frac{dy}
 {\sqrt{y^3+\frac{2A}{D}y^2 -4EDy + \frac{2A'}{D}\frac{L^2}{\hbar^{2}}}}.
\label{teqint}
\end{equation}
Note that L has the meaning of the third component of the angular
momentum in the space spanned by the coordinates $q_1,q_2,q_3$ with
$q_3$ a third coordinate which might be associated in a more complex
situation to an additional degree of freedom.

In our numerical 
application we considered two sets of parameters:
\begin{eqnarray}
I)~~ A'&=&0.01MeV, A = 3MeV, D=0.4MeV,\nonumber \\
II)~ A'&=&0.01 MeV, A = 3 MeV, D=-0.04 MeV. 
\label{IandII}
\end{eqnarray}
The two situations are represented pictorially in Figs. 2,3 for several 
values of L. One
notes that while the first effective potential has only one extreme value, 
a minimum, which depends on angular momentum, in the case labeled with II,  there
are two extremes, one minimum and one maximum. While the minima are depending
 on angular momenta, maxima are almost independent. Moreover there
is a critical value for angular momentum where the two extremes get unified
into an inflexion point. Below the critical angular momentum the effective
potential has a pocket shape and is similar to that obtained in the study
of heavy ion collision with two centers harmonic potential. 
Equation (\ref{toten}) suggests that the motion is allowed only for energies obeying the
restriction:
\begin{equation} 
E \ge V_{eff}^{min} 
\label{LgeV}
\end{equation}
For a given pair of (E,L) the intervals of $r$ were the
motion takes place are:
\begin{eqnarray}
r_{1} &\leq & r \leq r_{2},~~~\rm{ and}~~~ r \geq r_{3}~~\rm{ for}~~ E \leq 
V_{eff}^{max},\nonumber\\
r &\geq & r_{1} ~~\rm{for}~~ E \geq V_{eff}^{max},
\label{intervr}
\end{eqnarray}
were $r_{k}$ denotes the values of r were $E=V_{eff}$. The minimum and
maximum values of effective potential are denoted by $V_{eff}^{min}$ and
$V_{eff}^{max}$, respectively. Classical trajectories may evolve, 
depending on the initial conditions,
either on a finite trajectory in the interval $r_{1} \leq r \le r_{2}$ or on 
an unbound one with $r \geq r_{3}$. As we shall see, in the first interval the 
classical
motion could be quantized while in the second one the system undergoes a
fission process. In our semi-classical quantization procedure the system 
cannot go, through a tunneling effect, to the unstable region. However, in a 
full quantum mechanical description the wave function describing the system 
inside
the pocket region, is spread also to the region $r \geq r_{3}$.

Since t is a real quantity the integration interval in (\ref{teqint}) is chosen so that
inside it the polynomial is positive. As shown in Figs. 4 and 5 these
intervals are:
\begin{equation}
e_2\le x \le e_3 ~~{\rm for ~~case~~ I},~~e_1\le x \le e_2~~
{\rm~~ for~~ case~~ II}. 
\label{intervx}
\end{equation}
This suggests that a possible initial condition is:
\begin{eqnarray}
x_0\equiv x(0)=\left \{\matrix{e_2,&~~ {\rm~~ case~~ I}\cr
                              e_1.&~~ {\rm~~ case~~ II,~~ bound~~ trajectory}\cr
                              e_3,&~~ {\rm~~ case~~ II,~~ unbound~~ trajectory}}
\right \}
\label{initcon}
\end{eqnarray}

The integral involved in Eq. (\ref{teqint}) can be analytically performed
\cite{Rij}
and the final result for time is:
\begin{eqnarray}
t&=&\frac{\sqrt{2}\hbar}{\sqrt{|A'D|(e_3-e_1)}}sn^{-1}(u,k),
\nonumber\\
u&=&\sqrt{\frac{(e_3-e_1)(x-e_2)}{(e_3-e_2)(x-e_1)}}, 
~k=\sqrt{\frac{e_3-e_2}{e_3-e_1}},~~
{\rm for~~ the~~ case~~ I}\nonumber\\
u&=&\sqrt{\frac{x-e_1}{e_2-e_1}},~~k=\sqrt{\frac{e_2-e_1}{e_3-e_1}},~~
{\rm for~~ the~~ case~~ II},~e_1\le x\le e_2\nonumber\\
u&=&\sqrt{\frac{e_3-e_1}{x-e_1}},~~k=\sqrt{\frac{e_2-e_1}{e_3-e_1}},~~
{\rm for~~ the~~ case~~ II},~x\ge e_3.
\label{uandk}
\end{eqnarray}
Here $sn(u,k)$ denotes the elliptic  sinus function.  Its argument "u"
depends on 'x', the upper limit of the integral defining the time as
well as on the coordinates $e_1,e_2,e_3$ where the polynomial lying
under the square root symbol vanishes. Actually these are obtained by
squaring $r_1,r_2,r_3$ respectively, the turning points of the effective potential
mentioned above. The roots $e_k$ are ordered as follows:
\begin{equation}
e_1\le e_2 \le e_3.
\label{ordere}
\end{equation}
This function can be inverted
\begin{equation}
r^{2} = g(t),
\label{r2eqg} 
\end{equation}
and finally the coordinate r is expressed as a function of time.
The explicit expression for the coordinate x as a function of time is
given in Appendix A.
 Using this
result in connection with the second equation (2.13), the equation for the coordinate
$\theta$ is solved
\begin{equation}
\theta = \int_{0}^{t} \frac{A' L}{\hbar^{2} g(t')}dt' \equiv h(t)  
\label{theta}
\end{equation}
and thereby the classical trajectory is fully determined. 

An alternative way to derive the implicit equation of the classical 
trajectory is described in Appendix A.

For the variable $x$, the motion is periodical with the period:
\begin{equation}
T=\frac{\pi}{\sqrt{2|A'D|(e_3-e_1)}}{_2}F_1(\frac{1}{2},\frac{1}{2},1;k^2)
\label{period}
\end{equation}
where $_2F_1$ denotes the confluent hyper-geometric function with the argument
$k^2$, where $k$ is defined by Eq.(\ref{uandk}).

Note that the period depends on the energy $E$, by means of the roots
$e_1, e_2,e_3$.
Energy is discretitized through the quantization equation of
Bohr-Sommerfeld type:
\begin{equation}
{\cal I}(E)\equiv\frac{1}{2\pi \hbar}\int_{V^{min}_{eff}}^{E}T(E')dE'=n.
\label{IntofE}
\end{equation}
The function appearing in the left hand side of the above equations is
plotted in Figs. 6, 7 as function of E for the two particular
potentials  considered here. One remarks the fact that this integral
depends almost linearly on energy, irrespective of L. However the
slope of these straight lines depends on the potential characterizing the
motion of the system.

Quantized energies for n=0,2 and L=even and those with n=1 and $L\ge 2$
obtained for the pocket like potential are plotted in Fig. 8.
One should note that although these states describe the intrinsic
degrees of freedom, they might be organized in bands as it happens with
the states characterizing the whole set of degrees of freedom. 
In the case II the bands are finite since trajectories with energy larger
than $V^{max}_{eff}$ cannot be quantized, describing unbound systems.
Obviously the number of states which could appear inside the pocket
depends on angular momentum. The larger  angular momentum, the
smaller the number of the bound states. We mention again that in a pure
quantum mechanical treatment these states are quasi-bound.
In contrast with this situation, 
for the case I,  the bands are all infinite.

From Figs. 6-8, it results that for a given value of angular momentum
the energy spacings of states with different values for the quantum
number n is almost constant. 
This feature is determined by the relative
values of the coefficients defining the model Hamiltonian. 
For this particular situation, the harmonic approximation seems to be quite
reasonable. Indeed, expanding the effective potential around its
minimum and ignoring the terms of order higher than two, one finds for
the harmonic frequency the value:
\begin{equation}
\omega_L=\left [3A'\left (\frac{A'L}{r_0^4}+\frac{1}{3}A+Dr_0^2\right
)\right ]^{\frac{1}{2}}  .
\label{omegaL}
\end{equation}
Thus, the lowest three bands can be defined in an approximative way by:
\begin{eqnarray}
E^{(g)}_L&=&V^{min}_{eff}(L,r_0), L=0,2,4,6,...\nonumber\\
E^{(\gamma)}_L&=&V^{min}_{eff}(L,r_0)+\omega_L, L=2,3,4,...\nonumber\\
E^{(\beta)}_L&=&V^{min}_{eff}(L,r_0)+2\omega_L, L=0,2,4,6,...
\end{eqnarray}

Exact energies of the levels in the three bands are given in Fig. 8 
as function of angular momentum L for the case II.
From Fig. 7 one may see that the number of the bound states in the effective 
potential

characterizing the case II depends on the value of the angular momentum. 
This dependence is seen more
clearly in Fig. 9. From there one sees that the number of bound states varies 
from 18, for $L=0$, to 0, for $L\ge 46$.
Concluding this section, although the starting Hamiltonian is
anharmonic and has a complex structure we derived analytical solutions for the
classical trajectories. The corresponding energies where quantized using
a Bohr Sommerfeld type quantization condition. Energy levels are labeled by
two indices, n,L, one provided by the quantization procedure and the other
one by a conservation law. We fixed, conventionally, the constant
value of ${\cal L}_3$  to be equal to L/2. 
However the classical trajectory is not invariant to the fictitious
group $SU_c(2)$. Therefore it is deformed being a mixture of
components of different $L$. This can be easily seen if we evaluate
${\cal L}^2$ for a trajectory of energy E and projection of angular
momentum on z-axis equal to M:
\begin{eqnarray}
{\cal L}^2\equiv \langle \Psi|{\hat L}^2|\Psi\rangle&=&
\frac{1}{4A'}\left[E+(A'-A)r^2-\frac{D}{4}r^4\right ]
\nonumber\\
&+&
\frac{1}{4{A'}^2}\left[E+(A'-A)r^2-\frac{D}{4}r^4\right ]^2-
\frac{1}{2}\frac{M^2}{\hbar^2}.
\label{L2}
\end{eqnarray}
From this equation it is manifest that ${\cal L}^2$ is not a constant
of motion since this quantity depends on r and therefore changes its
value when the system moves from one point to another.

The problem treated in this Section is a real challenge for looking for
a quadrupole boson Hamiltonian which might yield in the intrinsic frame
a motion which admits both ${\cal L}_z$ and ${\cal L}^2$ 
as constants of motion. This would allow us to classify the intrinsic
motion with a ${\cal SU}(2)$ symmetry.
\section{Another solvable boson Hamiltonian}
\label{sec:level3}
Here we study the fourth order boson Hamiltonian:
\begin{equation}
H_2=\epsilon\sum_{\mu}b^{\dagger}_{\mu}b_{\mu}+\sum_{J=0,2,4}C_J
\left [(b^{\dagger}b^{\dagger})_J(bb)_J\right ]_0.
\label{H2}
\end{equation}
Along the time , this Hamiltonian has been used by several authors to describe
the rotational ground band. First authors were Das,  Dreizler and A. Klein 
\cite{Das} who treated this Hamiltonian in a particular basis and
obtained an analytical formula for the angular momentum dependence of
the yrast states energies. Later on this boson number conserving 
Hamiltonian was used by Iachello Ref.\cite{Iach,Ari}
to formulate the first version of the interacting boson approximation.
In \cite{Rad0}, this Hamiltonian was averaged on a angular momentum
projected state obtained from an axially deformed coherent boson
state to approximate the energies from a rotational ground band . 
In the vibrational limit the empirical formula of Ejiri \cite{Eji}
was rigorously derived.

As shown in Appendix B, the Hamiltonian $H_2$ can be written in an
equivalent otherwise more convenient form:
\begin{equation}
H_2=(\gamma+\epsilon){\hat N}_2+\beta{\hat N}^2_2+\delta{\hat J}^2_2+
\alpha(b^{\dagger}b^{\dagger})_0(bb)_0
\label{H2new}
\end{equation}
with the coefficients defined there.
The notation ${\hat N}_2$ is used for the quadrupole boson number
operator while ${\hat J}^2_2$ stands for the total angular momentum
squared carried by the quadrupole bosons.
The classical energy function associated to $H_2$ has the expression:
\begin{equation}
{\cal H}_2=A(u_0^2+v_0^2+2u_2^2+v_2^2))+B(u_0^2+v_0^2+2u_2^2+v_2^2))^2+
C(u_0v_2-u_2v_0)^2,
\label{H2uV}
\end{equation}
with the coefficients A, B, C given in Appendix B.

The classical equations of motion are obtained from the time dependent
variational equation (\ref{varpr}) by replacing $H_1$ with $H_2$. They have a
canonical form with respect to the coordinates $\{q_k,p_k\}_{k=1,2}$
defined by Eq. (\ref{qandp}). In terms of canonical coordinates the energy
function has the expression:
\begin{equation}
{\cal H}_2=\frac{A}{2}(q_1^2+q_2^2+p_1^2+p_2^2)+\frac{B}{4}
(q_1^2+q_2^2+p_1^2+p_2^2)^2+\frac{C}{8}(q_1p_2-q_2p_1)^2.
\label{H2clas}
\end{equation}
Since the boson Hamiltonian commutes with ${\hat L}_3$ and ${\hat N}_2$
there are two independent constants of motion:
\begin{equation}
{\cal N}_2=\langle\Psi|{\hat N}_2|\Psi\rangle,~~{\cal L}_3=\langle
\Psi|{\hat L}_3|\Psi \rangle
\label{N2andL3}
\end{equation}
This results from the equations:
\begin{eqnarray}
\frac{d}{dt}(q_1^2+q_2^2+p_1^2+p_2^2)&=&0,\nonumber\\
\frac{d}{dt}(q_1p_2-p_1q_2)&=&0,
\label{constmot}
\end{eqnarray}
implied by the equations of motion. Using these results one finds out
that ${\cal H}_2$ is a constant of motion.
On the other hand, considering the expressions of the $SU_c(2)$ generators
one finds that:
\begin{equation}
{\cal L}^2=\left[\frac{\hbar}{4}(q_1^2+q_2^2+p_1^2+p_2^2)\right]^2.
\label{L22}
\end{equation}
and according to the previous equation this is a constant of motion.
The classical energy can be quantized in two equivalent ways, namely either
one quantizes the angular momentum by fixing the constants of motion
such that:
\begin{equation}
{\cal L}^2=\hbar^2L(L+1),~~{\cal L}_3=\hbar M,~~{\rm with}~~-L\le M\le
L,~~L,M~~ {\rm integers}.
\label{quantL}
\end{equation}
or by quantizing the classical action
\begin{equation}
\int(q_1p_1+q_2p_2)dqdp=2\pi\hbar n, n-~{\rm positive~~ integer}
\label{pqquant}
\end{equation}
and the third component of angular momentum.
For the first option the result for the quantized energy is:
\begin{equation}
E_{LM}=2A\sqrt{L(L+1)}+4BL(L+1)+\frac{C}{2}M^2,
\label{ELM}
\end{equation}
while in the second case the result is:
\begin{equation}
E_{n,M}=A(n+1)+B(n+1)^2+\frac{C}{2}M^2
\label{EnM}
\end{equation}
In this equation we introduced the zero point motion for the plane
oscillator quanta although the semi-classical quantization in not able to
account for it.
To compare the two alternative expressions for the quantized energy it
is convenient to make the following approximations, which work quite well for
large quantum numbers:
\begin{equation}
(n+1)^2\approx n(n+2),~~\sqrt{L(L+1)}\approx L+\frac{1}{2}
\label{aprox}
\end{equation}
In this way the two expressions for  energy are identical provided
$L=\frac{n}{2}$.
It is worth mentioning that the Hamiltonian considered in this section 
was used in Ref. 17 in the 
boson basis
$\{|Nv\alpha JM\rangle\}$, 
where  the quantum numbers involved are the number of bosons, seniority, 
missing quantum number , angular momentum and its projection on z-axis, 
with the restriction $N=v=\frac{J}{2}$ for the ground band states.
By contrast, the connection between the new angular momentum quantum number 
L and the number of quanta in plane is
 $L=\frac{n}{2}$. 
This symmetry obeyed by the plane oscillator was exploited by Moszkowski
in Ref.\cite{Mosz} where a schematic solvable many body Hamiltonian was found which
describes fairly well the main features of the competition between
individual and collective degrees of freedom.

The equation (\ref{ELM}) shows that for the model Hamiltonian considered in
this section the energy of the intrinsic degrees of freedom can be
classified by the quantum numbers (L,M) and correspondingly the
quantized states by the irreducible representations of the fictitious 
$SU_b(2)$ group. Since energies depend on M, the states are deformed
despite the fact L is a good quantum number.
From the classical energy function it is clear that one deals with a
plane oscillator. For even values of "n" one may define  a one dimensional 
oscillator  whose 
number of quanta is determined as $2 \nu=n$.
Also we fix the projection of angular momentum on z-axis by the condition 
${\cal L}_3=\sqrt{L(L+1)}$.
Under these circumstances the quantized energy gets the expression:
\begin{equation}
E_{\nu,L}=2A(\nu+\frac{1}{2})+4B(\nu+\frac{1}{2})^2+\frac{C}{2}L(L+1).
\label{Eniu}
\end{equation}
Apart from an additive constant term this equation coincides with
those used by Erb and Bromley to fit the spectrum of $^{12}C+^{12}C$ system
\cite{Erb,Grei}. 

Using a more complex structure for the starting model
Hamiltonian, a coupling of rotational and vibrational degrees of
freedom is possible. Such a coupling has been described by
Iachello \cite{Ia81,Roos82,Lev82} in the framework of an
algebraic model.

Equations derived in this Section for energy refer to the intrinsic
degrees of freedom. Passing to the laboratory frame, the total energy
accounts also for the rotational degrees of freedom. Assuming that the
intrinsic degrees of freedom and Eulerian angles, defining the position 
of the intrinsic frame,
are only weakly coupled, the total energy can be approximated by:

\begin{equation}
E_{JLM}=2A\sqrt{L(L+1)}+4BL(L+1)+\frac{C}{2}M^2+\delta J(J+1).
\label{EJLM}
\end{equation}

Concluding this Section one may say that the energies of the exactly solvable
Hamiltonian can be classified by a SU(2)$\otimes$SU(2) symmetry, one factor
describing the motion of intrinsic degrees of freedom while the other one
taking care for the motion of Eulerian angles.

In the next section we consider a boson Hamiltonian to which it
corresponds a classical energy function which is similar to the
mean field underlying the model proposed by Moszkowski \cite{Mosz}.

\section{A quadrupole-quadrupole boson Hamiltonian.}
\label{sec:level4}
In this section we study a quadrupole boson Hamiltonian which, in the
intrinsic frame is closely related to the schematic Hamiltonian introduced by
Moszkowski\cite{Mosz}, long time ago. This Hamiltonian has been used by
several authors to test various many-body approaches\cite{Rad1,Marsh}. 
For the sake of completeness we
present first the main ideas underlying the Moszkowski model (MM).
\subsection{Brief review of Moszkowski model}
The MM model considers a system of nucleons moving in a mean field,
consisting in a
two dimensional oscillator potential and a spin-orbit term, and
interacting among themselves through a quadrupole-quadrupole interaction.
Let us consider first the one body Hamiltonian:
\begin{equation}
H_{sp}\equiv H_{ho}+H_{so}=\frac{1}{2M}(p_x^2+p_y^2)+\frac{M\omega_0^2}{2}
(x^2+y^2)-C\vec{l}\vec{s}.
\end{equation}
It can be checked that $H_{ho}$ commutes with the quasi-spin operators:
\begin{eqnarray}
t_x&=&\frac{1}{4}\left[M\omega_0(x^2-y^2)+\frac{1}{M\omega_0}(p_x^2-p_y^2)\right],
\nonumber\\
t_y&=&\frac{1}{2}\left[M\omega_0xy+\frac{1}{M\omega_0}p_xp_y\right],
\nonumber\\
t_z&=&\frac{1}{2}l_z=\frac{1}{2}(xp_y-yp_x).
\end{eqnarray}
and therefore its eigenstates can be classified by the irreducible
representation of the $SU(2)$ group generated by $\{t_k\}_{k=x,y,z}$:
\begin{eqnarray}
\vec{t}^2|Nm\sigma\rangle&=&\frac{N}{2}(\frac{N}{2}+1)|Nm\sigma\rangle\equiv
t(t+1)|Nm\sigma\rangle,
\nonumber\\
t_z|Nm\sigma\rangle&=&\frac{1}{2}m|Nm\sigma\rangle.
\end{eqnarray}
Here N denotes the total number of quanta and m the azimuthal quantum number.
The spin is perpendicular on the (x,y) plane, the two possible
orientations being specified by $\sigma (=\pm)$.
It is worthwhile to notice that in terms of stretched coordinates:
\begin{eqnarray}
\left(\matrix{x'\cr y'\cr z'}\right)=\sqrt{\frac{M\omega_0}{\hbar}}
\left(\matrix{x\cr y \cr z}\right )
\end{eqnarray}
the components of quasi-spin operator are formally identical with those of
angular momentum ${\cal L}_k$, defined by Eqs. (2.13), (2.14). However the two sets of operators act
on different spaces.

Let us consider a many body system moving in the mean field described
by $H_{sp}$ and interacting by the QQ force, which in plane acquires a
very simple form:
\begin{equation}
H_{QQ}=-\frac{1}{4}(T_+T_-+T_-T_+)=-\frac{X}{2}(T^2-T_z^2).
\end{equation}
where T denotes the operator acting on the many body states:
\begin{eqnarray}
T_{\mu}&=&T_{\mu}(+)+T_{\mu}(-),\nonumber\\
T_{\mu}(\sigma)&=&\sum_{m,m'}\langle Nm\sigma|t_{\mu}|Nm'\sigma\rangle
c^{\dagger}_{Nm\sigma}c_{Nm'\sigma}
\end{eqnarray}
One distinguishes two limiting cases. When C=0 one obtains the two
dimensional version of the Elliott model \cite{Elli} which is suitable
for describing the collective rotations and quadrupole vibrations of a
many body system. The other regime, when the long-range interaction is
missing (X=0), simulates the shell model description in realistic situations.
The intermediate situations can be covered by a smooth variation of the
two strength parameters $C$ and $X$.
Renormalizing the mean field due to the two body interaction one
obtains a single particle Hamiltonian which in terms of stretched
coordinates looks as:
\begin{equation}
H_{sp}^{\prime}=\frac{\omega}{2}({x'}^2+{y'}^2+{p_{x^{\prime}}}^2+
{p_{y^{\prime}}}^2)-\frac{X}{32}({x'}^2+{y'}^2+{p_{x^{\prime}}}^2+
{p_{y^{\prime}}}^2)^2+\frac{X}{8}(x'p_{y^{\prime}}-y'p_{x^{\prime}})^2.
\end{equation}
As we shall see, this expression for the renormalized mean field is very useful.
\subsection{Two body interaction for quadrupole bosons.}
In this subsection we shall treat semi-classically the following
quadrupole boson Hamiltonian:
\begin{equation}
H_3=\Omega \hat{N}_2-\frac{F}{4}\sum_{\mu}Q_{2\mu}Q_{2-\mu}(-)^{\mu},
\end{equation}
where $Q_{2\mu}$ denotes the quadrupole operators defined as:
\begin{equation}
Q_{2\mu}=\sqrt{6}\left(b^{\dagger}b\right)_{2\mu}.
\end{equation}
Averaging this boson Hamiltonian on the coherent state (\ref{Psi}) and
approximating the average for the two body term by
$\sum_{\mu}\langle Q_{2\mu}\rangle \langle Q_{2-\mu}(-)^{\mu}\rangle$, one
obtains the following expression for the classical energy function:
\begin{equation}
{\cal H}_3=\Omega(q_1^2+q_2^2+p_1^2+p_2^2)-\frac{3}{7}
F[(q_1^2+q_2^2+p_1^2+p_2^2)^2
-4(q_1p_2-q_2p_1)^2].
\end{equation}
Comparing the energy function with the mean field corresponding to
Moszkowski model we realize that they are identical if we accept the following
relationships between the strengths involved in the two Hamiltonians:
\begin{equation}
\Omega=\omega,~~\frac{3}{7}F=\frac{1}{32}X.
\end{equation}
Therefore the spectrum of the quantized intrinsic Hamiltonian is:
\begin{equation}
E_{LM}=2\Omega\sqrt{L(L+1)}-\frac{48}{7}F[L(L+1)-M^2].
\end{equation}
The competition between the individual and collective feature in
Moszkowski model corresponds to the interplay of harmonic and anharmonic
terms in the boson interacting model.

Again we pointed out the possibility to classify the states
describing the intrinsic degrees of freedom by the irreducible
representation of an SU(2) group. 
\section{Conclusions.}
\label{sec:level5}
Three boson Hamiltonians of complex structure otherwise exactly
solvable, were semi-classically treated.
The first is a boson number non-conserving Hamiltonian. Its classical
trajectories are analytically expressed in terms of elliptic functions.
Closed orbits are quantized and explicit expressions for discrete energies
are obtained. They are organized in rotational bands following the
traditional scheme of the  liquid drop model. The difference is that here the
states describe the motion of intrinsic degrees of freedom and moreover
the angular momentum components generate a rotation group acting in a
fictitious space.

The second Hamiltonian commutes with the quadrupole boson number
operator and
yields in a classical framework a spectrum which is classified by
angular momentum and its projection on z-axes. In the laboratory frame
the spectrum is classified by the symmetry SU(2)$\otimes$SU(2). The
similarity with the molecular spectrum used for studying the quasi-bound
states in $^{12}C+^{12}C$ system was pointed out.

The third Hamiltonian includes, as  a two-body boson interaction, the
separable quadrupole-quadrupole interaction. The classical image of
this Hamiltonian is similar to the schematic Hamiltonian introduced by 
Moszkowski to study the interplay of individual and collective degrees
of freedom. Its spectrum is also classified according to the symmetry
SU(2)$\otimes$SU(2). 

Concluding, the present paper points out a certain SU(2) symmetry which
allows for a complete description of the intrinsic dynamic variables,
$\beta$ and $\gamma$. In a subsequent publication we shall investigate
the question whether this symmetry is observed experimentally or not.
Our optimism is supported by the fact that the considered Hamiltonians
were already used in connection with different approaches to describe
some of available data.

\section{Appendix A}
\renewcommand{\theequation}{A.\arabic{equation}}
\setcounter{equation}{0}
\label{sec:level A}
From Eq (\ref{uandk}) one can obtain analytical expressions for the coordinate
x as function of time. 

When the effective potential has only one extremum, the case I, the
result is:
\begin{equation}
x(t)=e_1+\frac{(e_3-e_1)(e_2-e_1)}{(e_3-e_1)-(e_3-e_2)sn^2\left [
\sqrt{\frac{|A'D|}{2\hbar^2}(e_3-e_1)}t;\sqrt{\frac{e_3-e_2}{e_3-e_1}}\right]},
\end{equation}
As for the potential exhibiting a pocket structure, one obtains:
\begin{eqnarray}
x(t)&=&e_1+(e_2-e_1)sn^2\left 
[\sqrt{\frac{|A'D|}{2\hbar^2}(e_3-e_1)}t;\sqrt{\frac{e_1-e_1}{e_3-e_1}}.\right],
~e_1\le x\le e_2,\nonumber\\
x(t)&=&e_3+(e_3-e_2)\frac{sn^2\left [\sqrt{\frac{|A'D|}{2\hbar^2}(e_3-e_1)}t;
\sqrt{\frac{e_3-e_2}{e_3-e_1}}.\right]}
{cn^2 \left [\sqrt{\frac{|A'D|}{2\hbar^2}(e_3-e_1)}t;
\sqrt{\frac{e_3-e_2}{e_3-e_1}}.\right]},~ x\le e_3.
\end{eqnarray}

Now  we describe briefly how to obtain the implicit equation for the
classical trajectory.
Inserting the expression of the time derivatives given by Eqs. (\ref{toten}) 
and (\ref{H1eqE}) in the identity:
\begin{equation}
\frac{d\theta}{dt}=\frac{d\theta}{dr}\frac{dr}{dt}
\end{equation}
one obtains:
\begin{equation} 
\theta = \sqrt{\frac{A'L^2}{-2D\hbar^2}} \int_{x_{0}}^{x}
 \frac{dy}
 {y\sqrt{y^3+\frac{2A}{D}y^2 -4EDy + \frac{2A'}{D}\frac{L^2}{\hbar^{2}}}},
~x=r^2.
\end{equation}
The integral involved in the above equation can be analytically
performed and the final result is:
\begin{eqnarray}
\theta=\frac{L}{\hbar}\sqrt{\frac{2|A'|}{(e_3-e_1)|D|}}\left \{
\matrix{-\frac{1}{e_1}\Pi\left(arcsin(u),k^2\frac{e_1}{e_2},k\right)+
\frac{e_2}{e_2-e_1}
sn^{-1}(u,k),~{\rm case ~~I,~~}e_2\le x \le e_3\cr
\Pi\left(arcsin(u),\frac{e_1-e_2}{e_1},k\right),~{\rm 
case~~II,~~}e_1\le x \le e_2 \cr
\Pi\left(arcsin(u),\frac{e_2}{e_3-e_2},k\right)+
\frac{e_3\sqrt{e_3-e_1}-\sqrt{e_3-e_2}}{e_2e_3\sqrt{e_3-e_1}}sn^{-1}(u,k),
~{\rm case~~II},x\ge e_3}\right \}
\end{eqnarray}
The arguments $u$ and $k$ depend on the interval and are those defined
by Eq. (2.29). We denoted by $\Pi$ the elliptic function of third rank
defined as:
\begin{equation}
\Pi(\phi,n,k)=\int_{0}^{\phi}\frac{d\alpha}{(1+nsin^2\alpha)
\sqrt{1-k^2sin^2\alpha}}.
\end{equation}.

\section{Appendix B}
\renewcommand{\theequation}{B.\arabic{equation}}
\setcounter{equation}{0}
\label{sec:level B}
The anharmonic terms involved in Eq(\ref{H2}) defining the Hamiltonian $H_2$
can be easily expressed in terms of boson number and total angular
momentum operators:
\begin{eqnarray}
\left [(b^{\dagger}b^{\dagger})_2(bb)_2\right ]_0&=&-\frac{7}{5}\sqrt{5}
{b^\dagger}b^{\dagger})_0(bb)_0
+\frac{4}{35}\sqrt{5}{\hat N}_2({\hat
N}_2-1)-\frac{1}{35}\sqrt{5}({\hat J}_2-6{\hat N}_2),\nonumber\\
\left [(b^{\dagger}b^{\dagger})_4(bb)_4\right ]_0&=&\frac{1}{7}
(b^{\dagger}b^{\dagger})_0(bb)_0
+\frac{1}{7}{\hat N}_2({\hat
N}_2-1)+\frac{1}{21}({\hat J}_2-6{\hat N}_2).
\end{eqnarray}
Using these expressions in connection with Eq. (\ref{H2}) one obtains $H_2$
in the form given by (\ref{H2new}) with:
\begin{eqnarray}
\alpha&=&C_0-\frac{2}{7}\sqrt{5}C_2+\frac{1}{7}C_4,\nonumber\\
\beta&=&\frac{4}{35}\sqrt{5}C_2+\frac{1}{7}C_4,\nonumber\\
\gamma&=&\frac{2}{35}\sqrt{5}C_2-\frac{3}{7}C_4,\nonumber\\
\delta&=&-\frac{1}{6}(\beta+\gamma).
\end{eqnarray}
Averaging $H_2$ on the coherent state (\ref{Psi}) one  obtains the classical
energy from Eq. (\ref{H2clas}) where the following notations have been used:
\begin{eqnarray}
A&=&\gamma+\epsilon+\beta+6\delta,\nonumber\\
B&=&\beta+\frac{\alpha}{5},\nonumber\\
C&=&-\frac{8\alpha}{5}.
\end{eqnarray}

\begin{figure}
\centerline{\psfig{figure=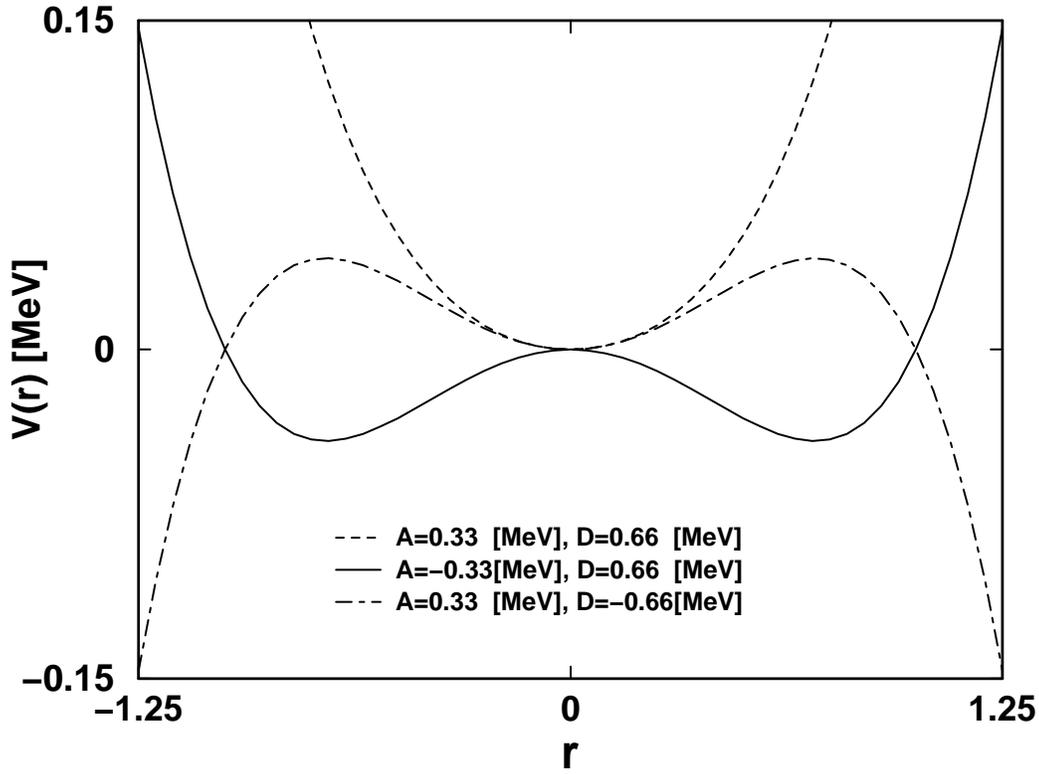,width=8cm,bbllx=5cm,%
bblly=7cm,bburx=18cm,bbury=26cm,angle=-90}}
\vspace*{4cm}
\caption{The potential energy involved in Eq. (2.11) is plotted as function
of r for three sets of parameters (A,D).}
\label{Fig. 1}
\end{figure}

\newpage
\begin{figure}
\centerline{\psfig{figure=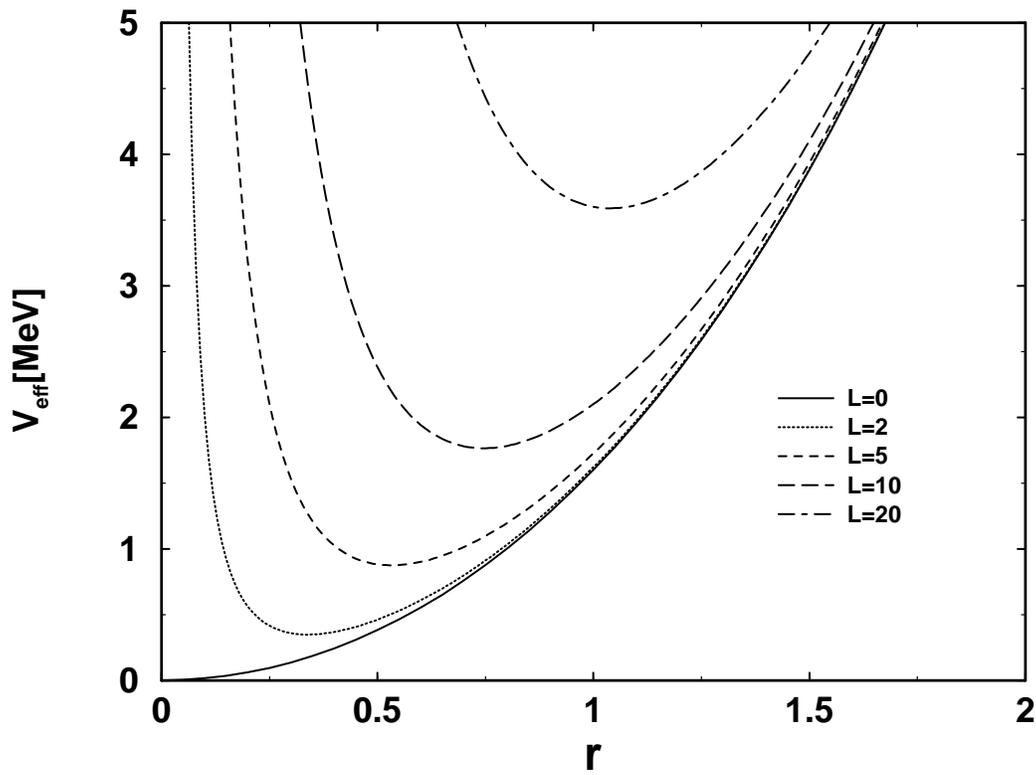,width=8cm,bbllx=5cm,%
bblly=7cm,bburx=18cm,bbury=26cm,angle=-90}}
\vspace*{4cm}
\caption{ The effective potential $V_{eff}(L,r)$ is plotted as
function of r for the parameters specified by Eq. (2.24) for the case I}.
\label{Fig. 2}
\end{figure}

\newpage
\begin{figure}
\centerline{\psfig{figure=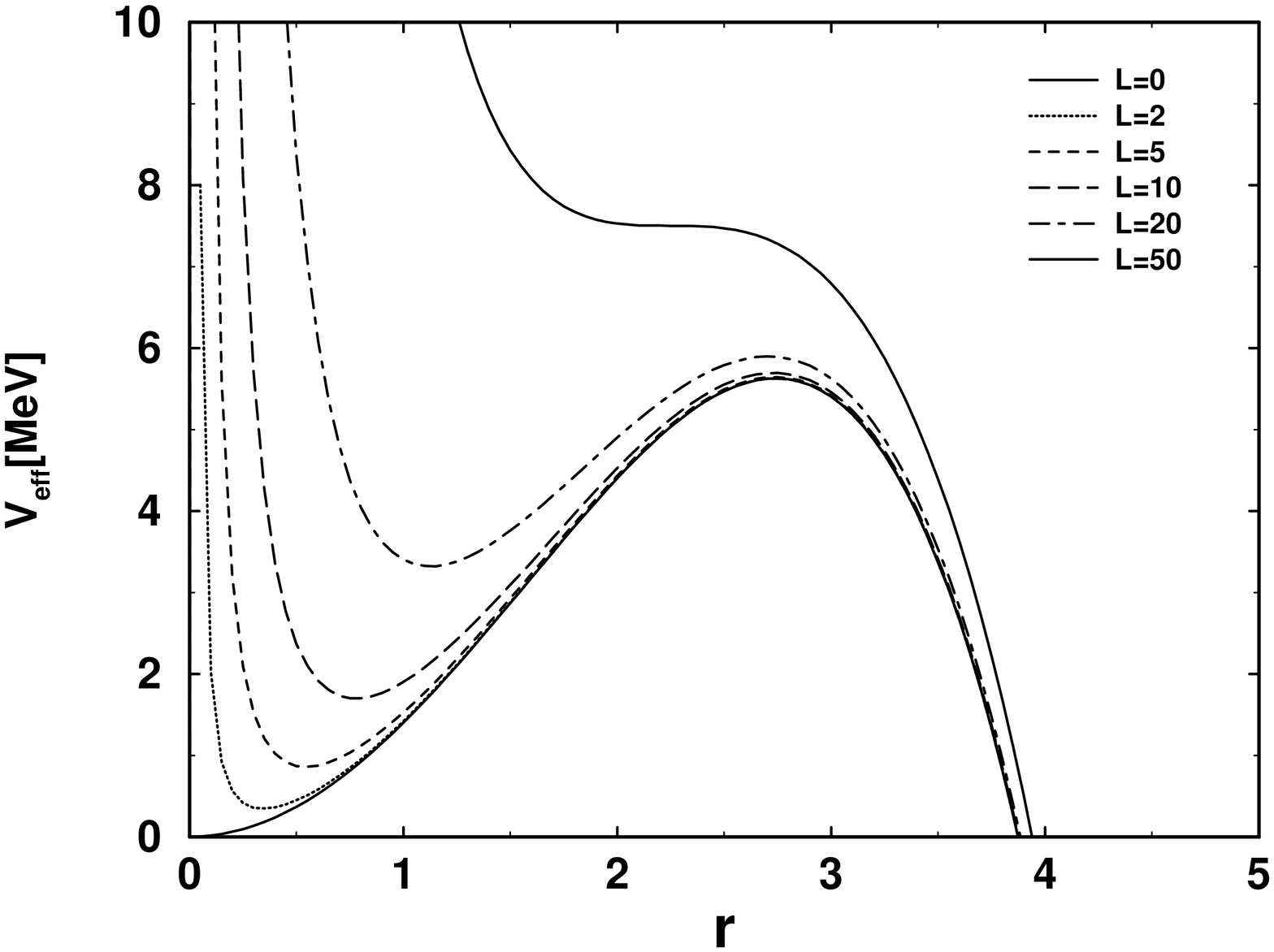,width=8cm,bbllx=5cm,%
bblly=7cm,bburx=18cm,bbury=26cm,angle=-90}}
\vspace*{4cm}
\caption{The same as in Fig. 2, but for the case II}
\label{Fig. 3}
\end{figure}

\newpage
\begin{figure}
\centerline{\psfig{figure=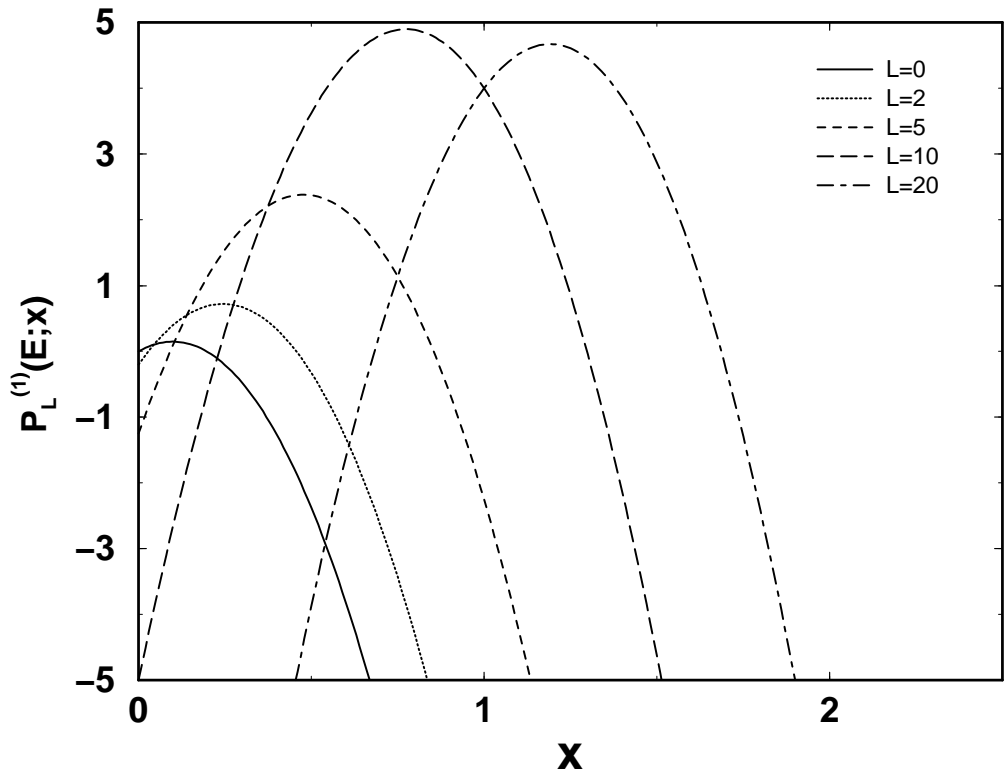,width=8cm,bbllx=5cm,%
bblly=7cm,bburx=18cm,bbury=26cm,angle=-90}}
\vspace*{4cm}
\caption{ The third order polynomial standing under the square root
symbol in Eq. (2.23), multiplied with $sign(-A'D)$ is plotted as
function of $x=r^2$, in the interval  $x\ge e_2$ where $e_1<e_2<e_3$
denotes its roots. The parameters correspond to the case I, listed in
Eq. (2.24). The plotted function is positive in the interval 
$e_2\le x\le e_3$.
}
\label{Fig. 4}
\end{figure}

\newpage
\begin{figure}
\centerline{\psfig{figure=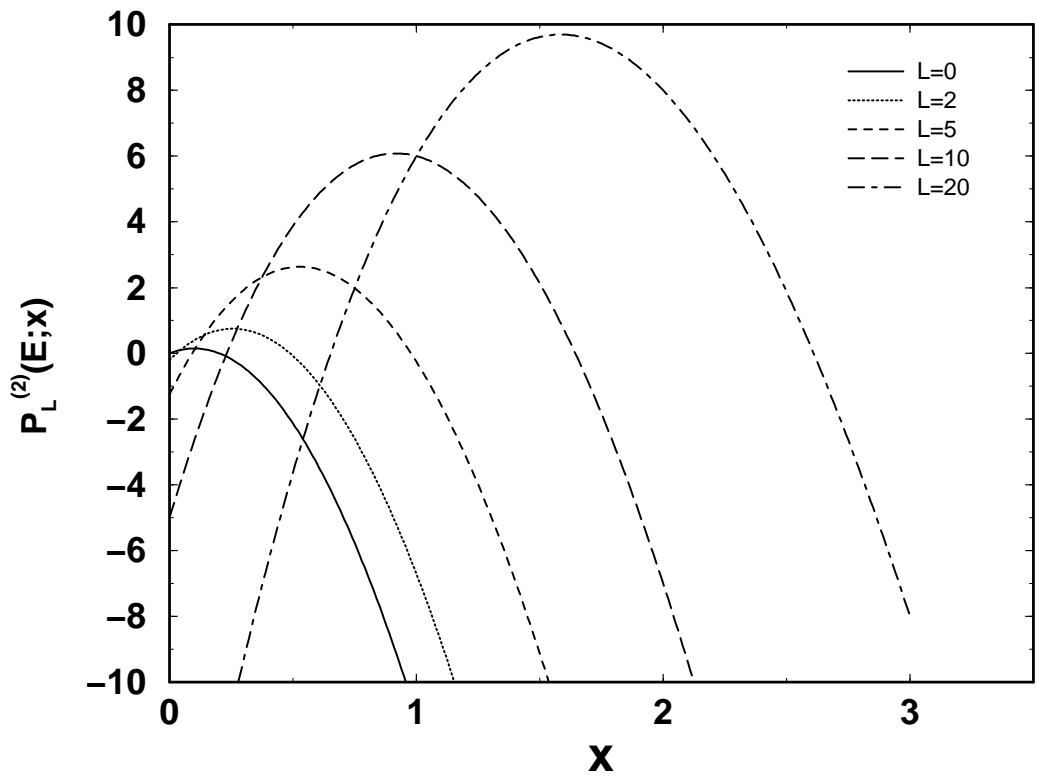,width=8cm,bbllx=5cm,%
bblly=7cm,bburx=18cm,bbury=26cm,angle=-90}}
\vspace*{4cm}\caption{ The same as in Fig. 4 but for the case II. 
The plotted
function is positive in the interval $e_1\le x \le e_1$
}
\label{Fig. 5}
\end{figure}

\newpage
\begin{figure}
\centerline{\psfig{figure=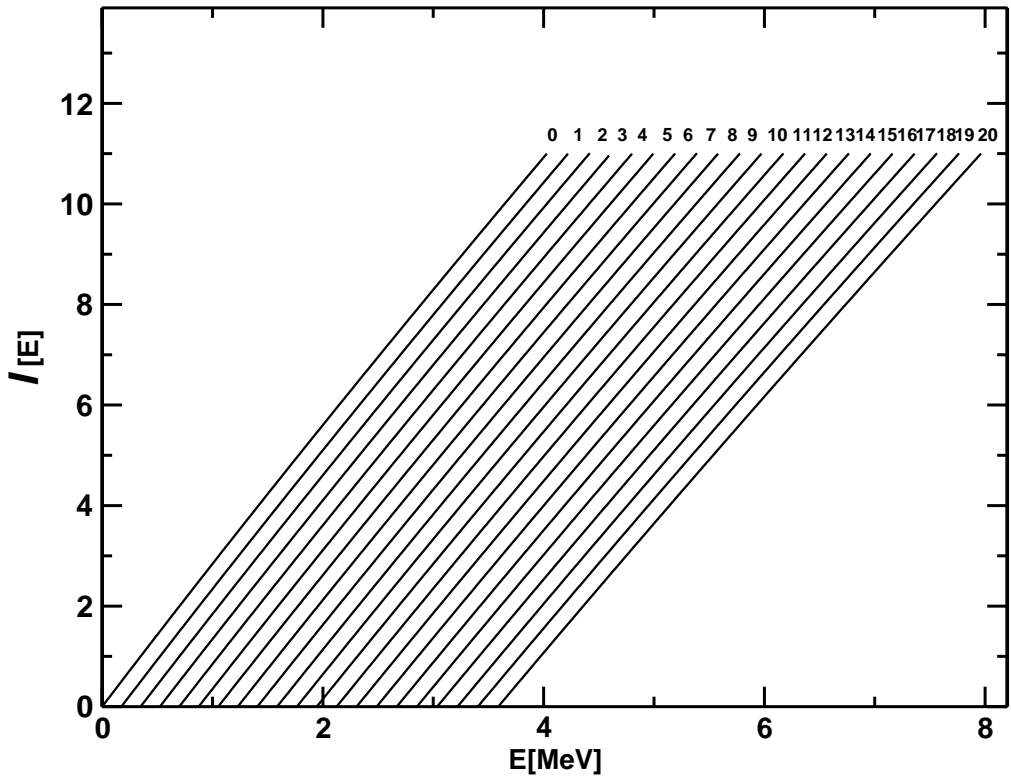,width=8cm,bbllx=5cm,%
bblly=7cm,bburx=18cm,bbury=26cm,angle=-90}}
\vspace*{5cm}
\caption{ The function ${\cal I}(E)$, defined by Eq. (2.34) is plotted
as function of E for several angular momenta. The parameters involved
are those specified in Eq. (2.24) by the label I.
}
\label{Fig. 6}
\end{figure}

\newpage
\begin{figure}
\centerline{\psfig{figure=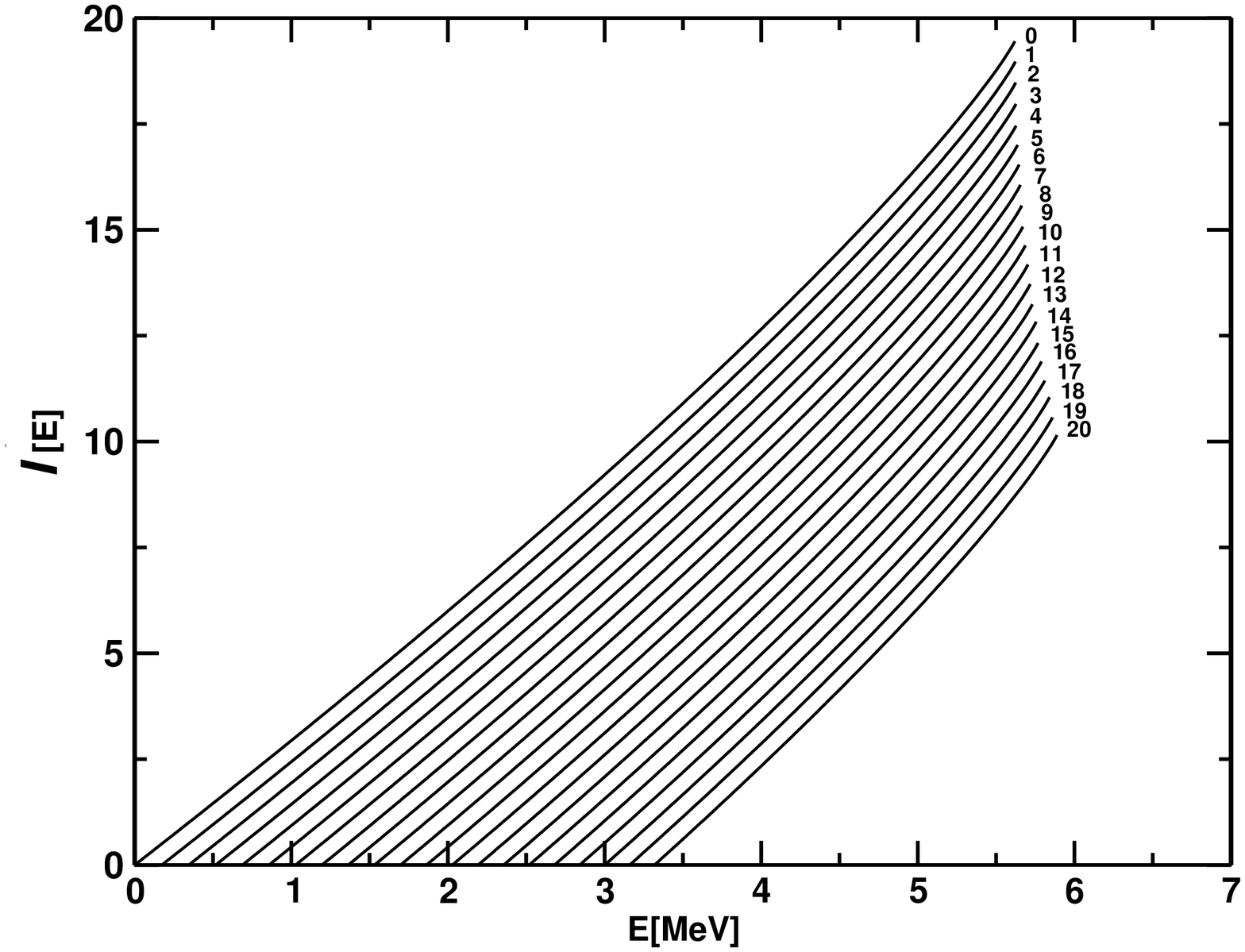,width=8cm,bbllx=5cm,%
bblly=7cm,bburx=18cm,bbury=26cm,angle=-90}}\vspace*{4cm}
\caption{The same as in Fig. 6, but for the case II}
\label{Fig. 7}
\end{figure}

\newpage
\begin{figure}
\centerline{\psfig{figure=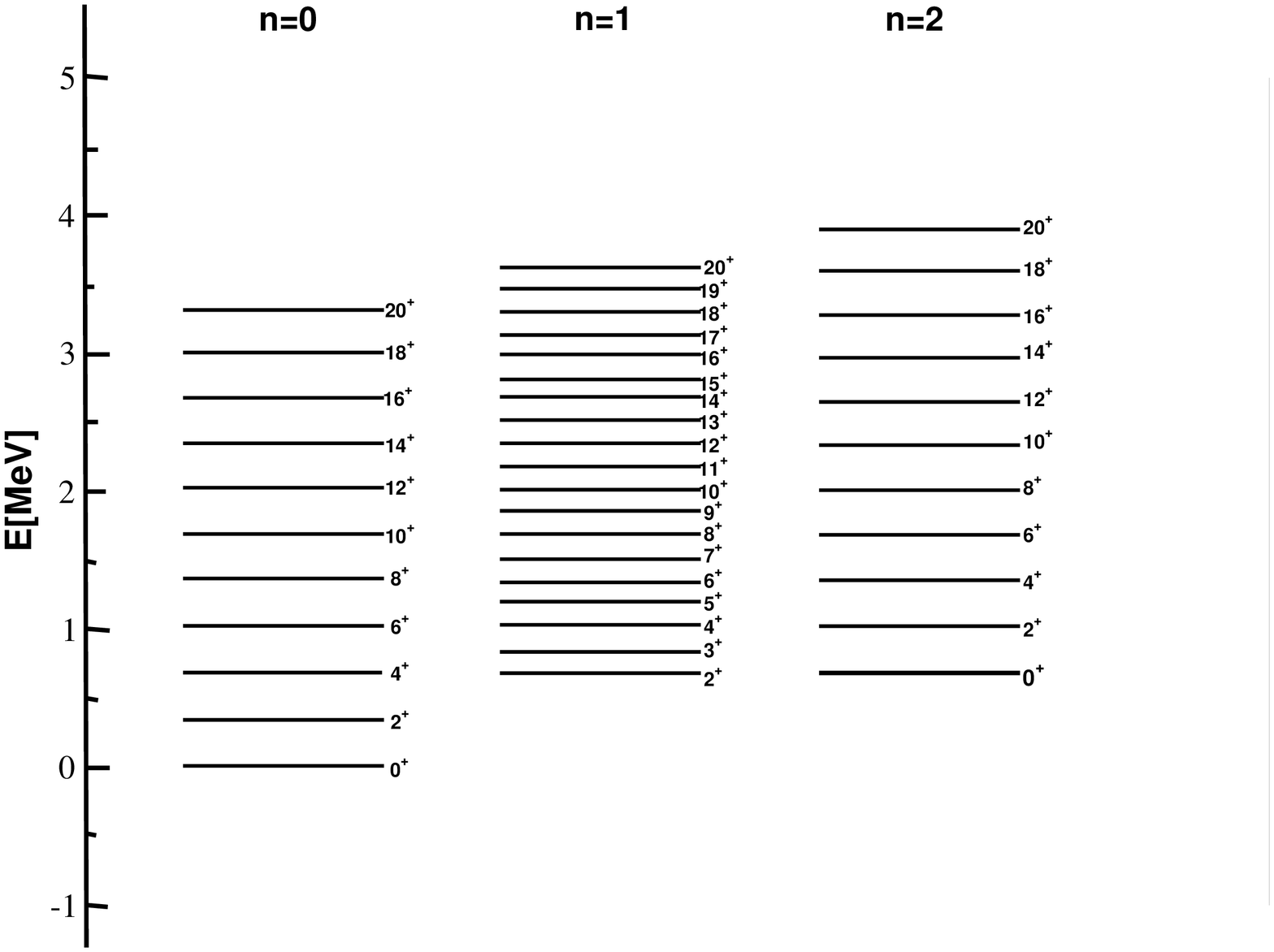,width=8cm,bbllx=5cm,%
bblly=7cm,bburx=18cm,bbury=26cm,angle=-90}}
\vspace*{4cm}\caption{ The energies associated to the motion of intrinsic degrees of
freedom are classified in rotational bands in the following way. The
lowest band is characterized by n=0 and even values for L. The second
band is similar to the gamma band and corresponds to n=1 and $L\ge 2$.
The third band is similar to the band $\beta$ and has n=2 and even
angular momenta.
}
\label{Fig. 8}
\end{figure}

\newpage
\begin{figure}
\centerline{\psfig{figure=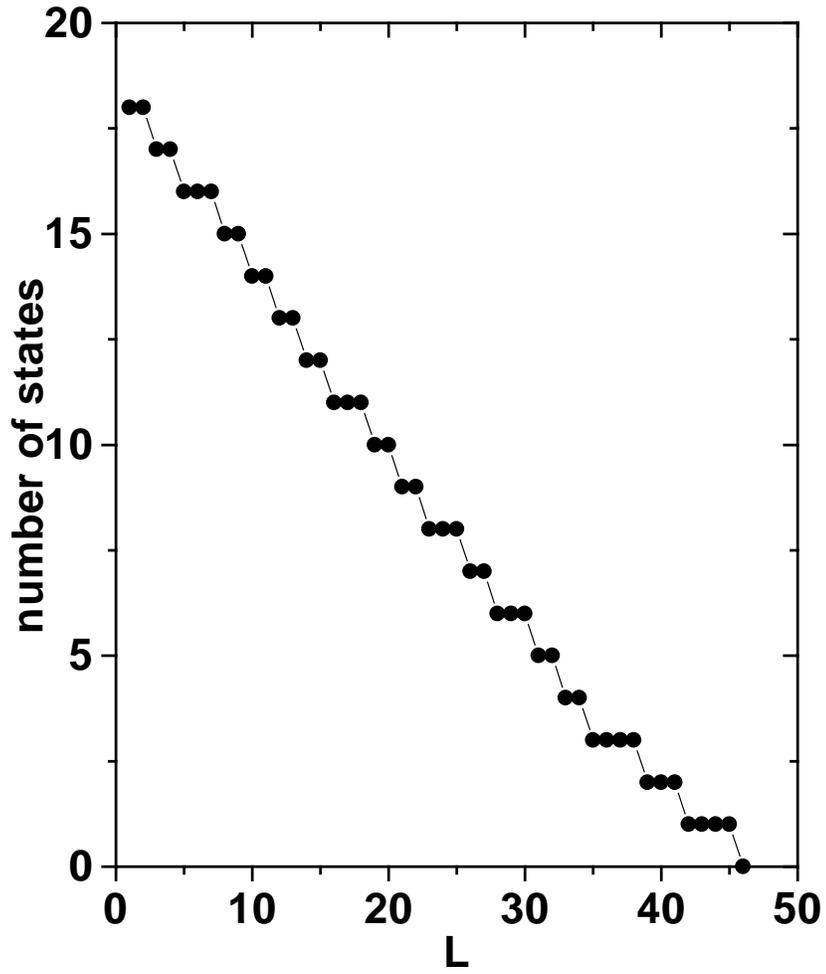,width=8cm,bbllx=5cm,%
bblly=7cm,bburx=18cm,bbury=26cm,angle=0}}
\vspace*{5cm}
\caption{The number of bound states in the pocket like potential
is plotted as function of angular momentum.
}
\label{Fig. 9}
\end{figure}
\end{document}